# Surface Composition of (99942) Apophis

Vishnu Reddy[1]
Lunar and Planetary Laboratory, University of Arizona, 1629 E University Blvd, Tucson, AZ 85721-0092
Email: reddy@lpl.arizona.edu

Juan A. Sanchez[1]
Planetary Science Institute, 1700 East Fort Lowell Road, Tucson, AZ 85719, USA

Roberto Furfaro
Systems and Industrial Engineering, University of Arizona, 1127 E. James E. Rogers Way, Tucson, AZ 85721-0020

Richard P. Binzel[1]
Department of Earth, Atmospheric, and Planetary Sciences, Massachusetts Institute of Technology, Cambridge, Massachusetts 02139

Thomas H. Burbine[1]
Department of Astronomy, Mount Holyoke College, South Hadley, MA 01075

Lucille Le Corre[1]
Planetary Science Institute, 1700 East Fort Lowell Road, Tucson, AZ 85719, USA

Paul S. Hardersen[1]
Planetary Science Institute, 1700 East Fort Lowell Road, Tucson, AZ 85719, USA

William F. Bottke
Southwest Research Institute, 1050 Walnut St, Suite 300, Boulder, CO 80302

Marina Brozovic
Jet Propulsion Laboratory, 4800 Oak Grove Drive, Mail Stop 301-120, Pasadena, CA 91109-8099

Pages: 33

Figures: 10
Tables: 2

**Proposed Running Head:** Apophis Composition

**Editorial correspondence to:**
Vishnu Reddy
Lunar and Planetary Laboratory,
University of Arizona,
1629 E University Blvd,
Tucson, AZ 85721-0092
Email: reddy@lpl.arizona.edu

**Abstract**

On April 13, 2029, near-Earth asteroid (NEA) (99942) Apophis will pass at a distance of ~6 Earth radii from Earth. This event will provide researchers with a unique opportunity to study the effects of tidal forces experienced by an asteroid during a close encounter with a terrestrial planet. Binzel et al. (2010) predicted that close flybys of terrestrial planets by NEAs would cause resurfacing of their regolith due to seismic shaking. In this work we present the best pre-encounter near-infrared spectra of Apophis obtained so far. These new data were obtained during the 2013 apparition using the NASA Infrared Telescope Facility (IRTF). We found that our spectral data is consistent with previous observations by Binzel et al. (2009) but with a much higher signal-to-noise ratio. Spectral band parameters were extracted from the spectra and were used to determine the composition of the asteroid. Using a naïve Bayes classifier, we computed the likelihood of Apophis being an LL chondrite to be >99% based on mol% of Fa vs. Fs. Using the same method, we estimated a probability of 89% for Apophis being an LL chondrite based on ol/(ol+px) and Fs. The results from the dynamical model indicate that the most likely source region for Apophis is the $\nu_6$ resonance in the inner main belt. Data presented in this study (especially Band I depth) could serve as a baseline to verify seismic shaking during the 2029 encounter.

# 1. Introduction

Aten-type potentially hazardous asteroid (PHA) (99942) Apophis has been the subject of an intense observational campaign since its discovery in 2004. This is because early estimates of the impact probability gave a chance of 1 in 37 for a collision with this object in the year 2029. As more observations were obtained and the orbit of the asteroid refined, the possibility of an impact was ruled out. Interest in this object, however, led to comprehensive multi-wavelength characterizations of this asteroid over the last decade.

Delbo et al. (2007) used polarimetric observations to estimate the absolute magnitude and albedo of Apophis, obtaining values of 19.7±0.4 and 0.33±0.08, respectively. These values correspond to a diameter of 270±60 m. Müller et al. (2014) revised these values by employing far-infrared observations obtained with the Herschel Space Observatory. Using a thermophysical model they derived a geometric albedo of 0.30 +0.05/-0.06, and a mean diameter of 375 +14/-10 m for Apophis. The thermal inertia was found to be in the range 250-800 J m$^{-2}$ s$^{-0.5}$ K$^{-1}$, with the best solution being $\Gamma$ = 600 J m$^{-2}$ s$^{-0.5}$ K$^{-1}$. According to Müller et al. (2014), these values would be compatible with a surface covered by a low conductivity fine regolith with rocks and boulders of high thermal inertia. They also noted that the best solution value is close to the one measured for Itokawa (700±200 J m$^{-2}$ s$^{-0.5}$ K$^{-1}$), which also has a very similar size, albedo, and interpreted composition.

More recently, Pravec et al. (2014) conducted a photometric campaign of Apophis and determined that it is in a non-principal axis rotation state ("tumbling"). The precession and rotation periods are 27.38 hours and 263 hours, respectively, with the strongest observed lightcurve amplitude for single axis mode with a 30.56 hours period. Apophis also turned out to be in retrograde rotation, which increased the probability for impact in 2068, but well below zero on the Palermo scale (Pravec et al. 2014). Pravec et al. (2014) also proposed a convex shape model for Apophis based on the lightcurve data.

A detailed mineralogical analysis of Apophis was performed by Binzel et al. (2009), who used the Shkuratov scattering model (Shkuratov et al. 1999) to model the spectra. They found that the best fit to Apophis spectrum is a mixture with an olivine-pyroxene abundance ratio (ol/(ol+px)) ranging from 0.65 to 0.75. Based on these results, Binzel et al. (2009) concluded that the best meteorite analog for Apophis were LL ordinary chondrites. This type of meteorite dominates the NEA population larger than 1 km (Vernazza et al. 2008; Dunn et al. 2013).

The close flyby of Apophis in 2029 presents us with a rare opportunity to observe a geophysical experiment. Previous studies (e.g., Binzel et al. 2010; DeMeo et al. 2014) have attributed the unweathered spectra of Q-type NEAs to seismic shaking during close planetary encounters that erases the spectral signatures of space weathering and refreshes the surface. Here we establish the "best" pre-encounter spectrum of Apophis, which was obtained during a favorable apparition on January 2013, when the asteroid was one and a half magnitude brighter than its previous apparition in 2005. The analysis of these new data differs from the work of Binzel et al. (2009) since mineral abundances are calculated using laboratory spectral calibrations that were not available by the time of their study. In addition, we also provide information about mafic silicate compositions for this asteroid.

## 2. Observations and Data Reduction

Near-IR observations of (99942) Apophis were obtained on January 14, 2013 using the SpeX instrument on NASA Infrared Telescope Facility (IRTF) on Mauna Kea, Hawai'i. Observational circumstances are presented in Table 1. Spectra of the asteroid, extinction, and solar analog stars were obtained in low-resolution prism mode (Rayner et al., 2003). Weather conditions were photometric throughout the observing window with an average atmospheric seeing of ~0.76" and a relative humidity of ~16%. All spectra were obtained at the parallactic angle to minimize differential refraction at the shorter wavelength end.

Twenty 120-second spectra of Apophis were obtained when the asteroid was 15.9 visual magnitude, at a phase angle of 41.8° and an airmass of ~1.37. Apart from Apophis, G-type local extinction star HD73877 was observed before and after the asteroid observations. The temporal and spatial proximity of the local G-type star observations enables better atmospheric modeling for correcting telluric bands. Thirty spectra of solar analog star HD 28099 were obtained to correct for spectral slope variations introduced by the use of a non-solar (i.e., G2V) local extinction star. Data reduction was performed using Spextool, a collection of IDL routines to perform wavelength calibration, telluric corrections, channel shifts, averaging, and display functions (Cushing et al., 2004). Detailed description of the data reduction procedure is presented in Reddy et al. (2009) and Sanchez et al. (2013).

## 3. Results

*3.1 Spectral characteristics of Apophis*

Figure 1 shows the average near-IR spectrum of Apophis normalized to unity at 1.5 µm. This spectrum exhibits two absorption features at ~1- and 2-µm, due to the presence of the minerals olivine and pyroxene. Binzel et al. (2009) observed Apophis on January 2005 when the asteroid was much fainter (17.4 V. Mag) using the SpeX instrument on NASA IRTF. The MIT-Hawaii Near-Earth Object Survey (MITHNEOS) also observed Apophis on January 2013 as part of their ongoing survey. Observational circumstances for data obtained by Binzel et al. (2009) and MITHNEOS are also included in Table 1. For comparison, we have plotted both spectra along with our spectrum in Fig. 2. The scatter seen in the spectra obtained by Binzel et al. (2009) and MITHNEOS at ~ 1.9 µm is primarily due to incomplete correction of telluric bands.

The taxonomic classification of Apophis was done using the online Bus-DeMeo taxonomy calculator (http://smass.mit.edu/busdemeoclass.html). We found that Apophis is classified as an Sq-type (PC1'=-0.1272, PC2'=0.0379) under this system (DeMeo et al. 2009), consistent with the classification given by Binzel et al. (2009).

Spectral band parameters including band centers, band depths and Band Area Ratio (BAR) were measured using a MATLAB code following the protocols described in Cloutis et al. (1986) and Gaffey et al. (2002). After removing the continuum, band centers were calculated by fitting a 2$^{nd}$ order polynomial over the bottom third of each band. Band areas, corresponding to the areas between the linear continuum and the data curve, were used to obtain the BAR, which is given by the ratio of area of Band II to that of Band I. Band depths were calculated using Eq. (32) from Clark and Roush (1984). The uncertainties associated with the band parameters are given by the standard deviation of the mean calculated from multiple measurements of each band parameter.

Temperature-induced spectral effects have been well documented (e.g., Singer and Roush, 1985; Moroz et al. 2000, Hinrichs and Lucey, 2002, Reddy et al. 2012a). With an increase or decrease in temperature, band centers can shift to longer or shorter wavelengths and absorption bands expand or contract. Correcting for these effects is an important step prior to mineralogical analysis. Therefore, we have calculated the average surface temperature of Apophis at the time of observation using Eq. (1) of Burbine et al. (2009), and applied the temperature correction of Sanchez et al. (2012) to the BAR value. Spectral band parameters are presented in Table 2.

*3.2 Compositional analysis*

Prior to the mineralogical characterization, spectral band parameters are plotted in the Band I center vs. Band Area Ratio plot to determine the S-asteroid subtype (Gaffey et al. 1993). Figure 3 shows the measured Band I center and BAR of Apophis together with the values measured for LL, L and H ordinary chondrites from Dunn et al. (2010). As can be seen in this figure, Apophis is located inside the polygonal region corresponding to the S(IV) subgroup of Gaffey et al. (1993). In particular, it lies in the LL ordinary chondrite zone, and just on the olivine-orthopyroxene mixing line of Cloutis et al. (1986). We have also measured the band parameters from the spectra obtained by Binzel et al. (2009) and MITHNEOS using the same procedure that we used with our data. These values are included in Table 2 and shown in Figure 3. The parameters extracted from both datasets also plot in the LL chondrite zone, slightly outside the polygonal region, which is likely attributed to the scattering of the data (see section 3.3.1).

The mineralogical characterization of Apophis was performed using the spectral calibrations derived by Dunn et al. (2010). These equations were derived from the analysis of ordinary chondrites and therefore can be used to accurately determine the surface composition of asteroids that fall in the S(IV) region. These equations make use of the Band I center to determine the olivine and pyroxene chemistry (given by the molar contents of fayalite (Fa) and ferrosilite (Fs)), and the BAR to calculate the abundance of these minerals in the assemblage (ol/(ol+px)). Reddy et al. (2014) used these spectral calibrations along with the band parameters measured for asteroid (25143) Itokawa, and demonstrated that mafic silicate compositions determined using this procedure are in excellent agreement with those measured from returned samples. These results give us confidence on the robustness of this technique when applied to an olivine-pyroxene assemblage like Apophis.

We applied Dunn et al. (2010) calibration and derived olivine and pyroxene chemistries of Apophis to be $Fa_{28.6}$ and $Fs_{23.6}$, respectively. These values, presented in Table 2, are consistent with the range for LL ordinary chondrites ($Fa_{25-33}$ and $Fs_{21-27}$) found by Dunn et al. (2010). In Fig. 4 we plot the molar content of Fa vs. Fs for Apophis. This figure also shows the calculated chemistries from Binzel et al. (2009) and the MITHNEOS data, and measured values for LL, L, and H ordinary chondrites from Nakamura et al. (2011). Binzel et al. (2009) values for Fa, and Fs, and those calculated from the MITHNEOS data are also consistent with those measured for LL chondrites. A quantitative analysis has been executed to estimate, given the data, the posterior probability of Apophis be in one of the chondrite classes. More specifically, a naïve Bayes classifier (see Appendix A.1) has been constructed to compute the likelihood of the derived mol% of fayalite (Fa) vs. ferrosilite (Fs) for Apophis [this work and also Binzel et al. (20090 and MITHNEOS data] to fall under H, L or LL ordinary chondrites classes. The measured values of H/L/LL ordinary chondrites (Figure 4) have been employed as training points to model the individual class likelihood and construct a discriminative classifier based on maximum a-posterior probability. The probability of the three classes (Blue = H, Green = L and Red = LL) is reported in Figures 5 and 6. More specifically, Figure 5 shows the decision boundaries estimated by the Bayes classifiers. Figure 6 (top) shows, for each value of Fa and Fs in the selected range, the computed probability distribution for each of the classes (i.e., H, L and LL). Figure 6 (bottom) also reports the contour plot that shows the value of the maximum a-posterior probability. For each value of Fa and Fs, the maximum probability computed by the Bayes classifier is plotted according to a color code. Binzel et al. (2009) measured value (black diamond) is located at the center of the data-driven computed distribution of LL ordinary chondrites whereas the values determined for this work (pink diamond) and MITHNEOS data (yellow diamond) fall in the lower part of this region. The posteriori likelihood of the Binzel et al. (2009) data is computed to be 99.9% LL and 0.1% L. The posterior likelihood of this work is computed to be 99.8% LL and 0.2% L, and that of the MITHNEOS data 99.5% LL and 0.5% L. The data points for this and previously reported measurements show that they fall well within the LL decision boundaries with respect to the probabilistic distribution of the three classes modeled using the data available from the Dunn database. The results should be interpreted as high (probabilistic) confidence of belonging to class LL. As a word of caution, the predicted posterior probability is reported "as computed" by the classifier given the limited amount of training points employed to model the distribution. Nevertheless, the results represent the best predicted probabilistic classification given the knowledge encoded in the training point distribution.

The ol/(ol+px) ratio for Apophis was found to be 0.63±0.03. Figures 7 shows the ol/(ol+px) ratio vs. mol% of Fs. Values found in this work fall in the region corresponding to LL ordinary chondrites. An ol/(ol+px) ratio of 0.60±0.03 was obtained from the Binzel et al. (2009) data, while a value of 0.66±0.03 was determined from the MITHNEOS data, in both cases falling within the range of LL chondrites. The same methodology described above has been applied to this case, i.e. deriving the maximum likelihood of belonging to class H, L or LL as function of ol/(ol+px) and Fs (Figures 8 and 9). We found that Binzel et al. (2009) data has a probability of being class LL as 98%, and a probability of being class L as 2%. Data presented in this work has a probability of being class LL as 89%, and a probability of being class L as 11%. The analysis of the MITHNEOS data indicates that Apophis has probabilities of 86% and 14% of being class LL and L, respectively.

*3.3 Comparison with previous work*

The analysis of these new NIR data shows that the spectrum of Apophis has a Band I center of 0.99±0.01 μm and BAR of 0.39±0.03. These values, however, differ from those obtained by Binzel et al. (2009), who measured a Band I center of ~ 1.055 μm and BAR of ~ 0.59 from their data. The ol/(ol+px) ratio calculated for the new data also differs from the value determined by Binzel et al. (2009). These differences could be the result of different factors, including: the procedure used to measure the band parameters, surface variegation (due to differences in composition, differences in grain size, or exogenic contaminants), and the analysis protocols. Here we explore each of these options to explain the observed differences between our data and that obtained by Binzel et al. (2009) and MITHNEOS.

*3.3.1 Procedure used to measure band parameters*

The first possibility to try to explain the observed differences is the use of different procedures to measure the spectral band parameters. In our case, we measure the Band I center after removing the continuum by fitting a polynomial over the bottom of the band. Band areas, used to determine the BAR, are measured using trapezoidal numerical integration. In contrast, Binzel et al. (2009) used the Modified Gaussian

Model (MGM) for measuring these band parameters. In their method, the Band I center is defined as the wavelength at which the band area is bisected by a vertical line. In addition, we have applied a temperature correction to the BAR, which was not available at the time the work of Binzel et al. (2009) was done. Our own measurements taken from the spectrum of Binzel et al. (2009) give values of 1.01±0.01μm and 0.54±0.07 for the Band I center and BAR, respectively. In the case of the data obtained by MITHNEOS, we found that the Band I center is 0.99±0.01 μm, while the BAR is 0.27±0.03. The difference in BAR could be attributed in part to the fact that the spectrum obtained in the present work extends to 2.5 μm, while the other two spectra only have useful data until 2.45 μm. The scattering beyond ~1.6 μm (Figure 2) seen in the spectra could also explain the differences in BAR, as this parameter is particularly sensitive to the point-to-point scatter of the data. Finally, the temperature correction applied to the BAR (see section 3.1) also produces a small decrease of this parameter.

*3.3.2 Surface Variegation*

The differences in band parameters and composition seen between our data and that obtained by Binzel et al. (2009) and MITHNEOS could be the result of compositional variations across the surface of the asteroid. We used the Pravec et al. (2014) shape model of Apophis from lightcurve inversion technique to identify the orientation of the asteroid when our observations were made. Figure 10A shows the plane of sky images of Apophis at January 14, 2013 at 11:07:41 UTC (asteroid centric) at a rotation phase of 189.4° and phase angle of 41.8° when we made the observations. Figure 10B shows the same but for January 17, 2013 at 10:59:11 UTC at a rotation phase of 5.6° when the MITHNEOS data were collected. The X indicates the location of the Sub-Earth point and the vertical arrow is shows the spin vector. The uncertainties in the spin state prevented us from accurately constraining the rotation phase/shape model for 2005 observations and hence it is omitted. However, using this information we can at least compare our data with the one obtained by MITHNEOS. As can be seen in this Figure, Apophis was in a completely different orientation on these two days; i.e., we were looking close to the opposite ends of the long axis. Thus, the observation of two different regions on the surface of the asteroid could explain the variations seen between both datasets. It is worth mentioning, however, that surface variegation has been only observed and confirmed on only large main belt asteroids such as Vesta so far (e.g., Reddy et al. 2012b). Even these color and compositional variations are related to in fall of exogenic material rather than endogenic causes. Near-earth asteroid Itokawa is the only asteroid in the size range of Apophis that has been visited by a spacecraft. Observations of Itokawa by the Hayabusa spacecraft (Ishiguro et al. 2010) have shown that the surface color and composition is homogenous except for one black boulder. However, the spacecraft observed a wide range in surface texture from boulders to fine regolith. These observations suggest that any surface spectral variations on small NEAs could be a particle size effect rather than just composition.

*3.3.3 Analysis Protocols*

The other minor difference between our results and those reported by Binzel et al. (2009) corresponds to the ol/(ol+px) ratio estimated for Apophis. Using the calculated BAR along with the equation derived by Dunn et al. (2010), we estimated the olivine abundance for Apophis in 63±3% (this work), 60±3% (Binzel et al. 2009 data), and 66±3% (MITHNEOS data). Binzel et al. (2009), on the other hand, obtained a value of 70±5%. The study of Binzel et al. (2009) came out one year before the empirical equations of Dunn et al. (2010) were published. In their work, Binzel et al. (2009) used the Shkuratov scattering model (Shkuratov et al. 1999) to derive the mineral abundances of Apophis. Vernazza et al. (2008) used the same model to estimate the average ol/(ol+px) ratio of ordinary chondrites, obtaining values of 75% for LL, 64% for L and 59% for H chondrites. These values are much higher than those obtained by Dunn et al. (2010) using their empirical equations (63% for LL, 57% for L and 52% for H chondrites). Thus, this difference between the two techniques could account for the different ol/(ol+px) ratios calculated for Apophis.

*3.4 Source region*

We used the model of Bottke et al. (2002) to determine the possible origin of Apophis. This model considers five possible source regions: the $\nu_6$ secular resonance, the Mars-crossing region, the 3:1 mean motion resonance with Jupiter, the outer belt region, and the Jupiter Family Comet region. The results from the dynamical model indicate that the most likely source region for Apophis is the inner main belt, with a probability of 59% that the asteroid originated in the $\nu_6$ resonance. Similar results were obtained using the model of Granvik et al. (2016), which gives a probability of 84% that Apophis derived from the $\nu_6$ resonance.

Due to their compositional affinity, LL chondrites have been associated with the Flora family (Vernazza et al. 2008, de León et al. 2010, Dunn et al. 2013; Reddy et al. 2014), which is located in the inner part of the main belt at ~ 2.3 AU from the Sun. Hence, NEAs exhibiting LL chondrite-like compositions are thought to have originated in the Flora family and delivered to the near-Earth space via the $\nu_6$ secular resonance (Nesvorný et al. 2002). Thus, if Apophis is an LL chondrite, it could have originated in this asteroid family.

*3.5 Apophis' close encounter in 2029*

On April 13, 2029, Apophis will pass at a distance of 35900±8980 km (~6 Earth radii) from Earth (Sheeres et al. 2005), providing a unique opportunity to study the effects of tidal forces on an asteroid during a close encounter with a terrestrial planet. Binzel et al. (2010) showed that tidal stress caused during close encounters with the Earth (within ~16 Earth radii) would produce landslides exposing fresh unweathered material. Furthermore, numerical simulations carried out by Scheeres et al. (2005) indicated that terrestrial torques would significantly alter Apophis' spin state during this close encounter. They speculated that this could result in localized shifts on the asteroid's surface. While it is not clear how extensive this resurfacing will be, if it occurs on a global scale it might be possible to detect it using ground-based telescopes. Spectrally, this surface refreshing would be seeing as a decrease in spectral slope and an increase in band depths, with the compositional interpretation remaining the same (Gaffey 2010).

As for the taxonomic classification, the principal components PC1' and PC2' would move from the Sq towards the Q-types in this parameter space. This is because for asteroids having an ordinary chondrite-like composition, Q-, Sq-, and S-types are thought to represent a weathering gradient, where Q-types have relatively fresh surfaces, and Sq- and S-types have increasingly more space-weathered surfaces (e.g., Binzel et al. 2001; Binzel et al. 2010). As an example, we measured the band parameters for the mean spectrum of a Q-type asteroid from DeMeo et al. (2009). We found that the Band I center (0.99±0.01 μm) has the same value measured for Apophis, while the Band I depth (23.8±0.01) shows an increment of 6.8 compared to the Band I depth measured for Apophis (17.0±0.01). Thus, this parameter could be used to identify fresh exposed material on the surface, as has been used in the past with the Moon and other asteroids (e.g., Lucey et al. 2000; Murchie et al. 2001; Shestopalov 2002; Golubeva and Shestopalov 2003).

## 4. Summary

Apophis is one of the most interesting near-Earth asteroids due to its close encounter to Earth in 2029 that would enable us to observe a live geophysics experiment. Our spectroscopic observations made during the 2012-2013 Earth flyby reveals the following:
- Our spectral data are consistent with previous observations by Binzel et al. (2009) except that our data was obtained when the asteroid was brighter (15.9 V. Mag) and hence has a higher signal to noise ratio. This enabled us to perform detailed mineralogical analysis using calibrations not available to Binzel et al. (2009).
- We applied Dunn et al. (2010) calibration and derived olivine and pyroxene chemistries of Apophis to be $Fa_{28.6}$ and $Fs_{23.6}$, respectively. These values are consistent with the range for LL ordinary chondrites ($Fa_{25-33}$ and $Fs_{21-27}$).

- We estimate the olivine abundance for Apophis to be 63±3 vol. %, which is also consistent with the value estimated for LL chondrites (Dunn et al. 2010).
- A naïve Bayes classifier was constructed to compute the likelihood of the derived Mol % of Fa vs. Fs for Apophis to fall under H, LL or L ordinary chondrites classes. The posterior likelihood of Apophis is computed to be >99% LL and <1% L. The same procedure was applied to derive the maximum likelihood of belonging to each class of ordinary chondrites as a function of ol/(ol+px) and Fs. In this case, we found that Apophis has a probability of being class LL of 89%, and a probability of being class L of 11%.
- The results from the dynamical model indicate that Apophis originated in the $\nu_6$ resonance, possibly from the Flora asteroid family located in the inner part of the main belt.
- The band parameters presented in this work, in particular the Band I depth, could serve as a baseline to verify seismic shaking during the 2029 encounter.

**Acknowledgment**

This research work was supported by NASA Solar System Observations Grant NNX14AL06G (PI: Reddy) and the Czech **G**rant Agency (grant P209-13-01308S). We thank the IRTF TAC for awarding time to this project, and to the IRTF TOs and MKSS staff for their support. All (or part) of the data utilized in this publication were obtained and made available by the MIT-UH-IRTF Joint Campaign for NEO Reconnaissance. We thank the anonymous reviewer for useful comments that helped improve this paper. Taxonomic type results presented in this work were determined, in whole or in part, using a Bus-DeMeo Taxonomy Classification Web tool by Stephen M. Slivan, developed at MIT with the support of National Science Foundation Grant 0506716 and NASA Grant NAG5-12355. THB would like to thank the Remote, In Situ, and Synchrotron Studies for Science and Exploration (RIS$^4$E) Solar System Exploration Research Virtual Institute (SSERVI) for support. Any opinions, findings, and conclusions or recommendations expressed in this material are those of the author(s) and do not necessarily reflect the views of NASA or the National Science Foundation.

**Appendix**

*A.1 Kernel Density Methods for Classification: Naïve Bayes Classifier*

Kernel density-based classification is an unsupervised learning technique that naturally leads to the design and implementation of a family of methods for non-parametric classification. Indeed, given a set of training data, one can use Bayes theorem to predict the probability of a new (unseen) data point to belong to one of the class. Given a random variable $X$, for an *I-class* problem, where $I$ is the number of classes, one can separately fit non-parametric density estimates $p_i(X)$, $i=1,..., I$ for each individual class. Given class priors $\pi_i$, i.e. prior probability of being in class $i$, one can compute the probability of class $i$ given the new data sample $x_o$:

$$p(class = i | X = x_0) = \frac{\pi_i p_i(x_0)}{\sum_{j=1}^{I} \pi_j p_j(x_0)}$$

Within the kernel density approach, the Naïve Bayes classifier applies the density estimation method to the available data. The naïve Bayes model assumes that for a given class $i$, the $M$ predictors (features) $x_k$, $k=1,...M$ are conditionally independent, i. e.:

$$p_i(x) = \prod_{k=1}^{M} p_{ik}(x_k)$$

Naturally, with this assumption, the estimation problem is drastically simplified because the individual class-conditional marginal densities $p_{ik}$ are estimated separately. The Naïve Bayes classifier assigns new data (i.e. observations) to the most probable class by computing the maximum a-posteriori probability (decision rule). For a set of M predictors, the posterior probability of class $i$ is computed as follows.:

$$p(class = i | x_1, x_2, ..., x_M) = \frac{\pi_i \prod_{k=1}^{M} p_{ik}(x_k)}{\sum_{j=1}^{I} \pi_j \prod_{k=1}^{M} p_{ik}(x_k)}$$

The method classifies an unseen data point by computing the posterior probability for each individual class and subsequently assign the new observation to the class that possesses the maximum posterior probability.

Although the assumption often tends to be violated for real data, in practice the naïve Bayes classifier yields posterior distributions that are robust to biased class density estimates. Indeed, despite the optimistic assumption of conditional independence of the predictors, naives Bayes classifiers tend to outperform kernel methods that are more sophisticated (Hastie et al, 2008).

**Tables**

**Table 1**: Observational circumstances. The columns in this table are: Date (UTC), V-magnitude, phase angle (⟨), and heliocentric distance (r). Observational circumstances corresponding to data obtained by Binzel et al. (2009) and MITHNEOS (file a099942.sp117.txt) are also included. All observations used the SpeX instrument on NASA Infrared Telescope Facility (IRTF) on Mauna Kea, Hawai'i.

| Observational circumstances | Date (UTC) | Mag. (V) | ⟨ (°) | r (au) |
|---|---|---|---|---|
| Binzel et al. (2009) | 01/08/ 2005 | 17.4 | 71.1° | 1.02 |
| This work | 01/14/ 2013 | 15.9 | 41.8° | 1.05 |
| MITHNEOS | 01/17/ 2013 | 15.8 | 38.0° | 1.06 |

**Table 2**: Spectral band parameters and composition for Apophis. Values measured from the spectrum obtained by Binzel et al. (2009), and MITHNEOS (file a099942.sp117.txt) are also presented. The errors corresponding to the olivine and pyroxene composition, and the ol/(ol+px) are given by the uncertainties derived by Dunn et al. (2010). The average surface temperature of Apophis was calculated as in Burbine et al. (2009), assuming a geometric albedo of 0.3 (Müller et al. 2014). Temperature corrections derived by Sanchez et al. (2012) were applied to the BAR and the ol/(ol+px).

| Parameter | This work | Binzel et al. (2009) | MITHNEOS |
|---|---|---|---|
| Band I Center (⌠m) | 0.99±0.01 | 1.01±0.01 | 0.99±0.01 |
| Band II Center (⌠m) | 1.87±0.04 | 1.84±0.11 | 2.03±0.08 |
| Band I Depth (%) | 17.0±0.1 | 15.3±0.3 | 17.2±0.2 |
| Band II Depth (%) | 6.8±0.1 | 3.3±0.3 | 5.2±0.2 |
| Band Area Ratio (BAR) | 0.42±0.03 | 0.56±0.07 | 0.30±0.03 |
| Temp. corrected BAR | 0.39±0.03 | 0.54±0.07 | 0.27±0.03 |
| Olivine composition (mol %) | Fa (28.6±1.3) | Fa (30.1±1.3) | Fa (28.4±1.3) |
| Pyroxene composition (mol %) | Fs (23.6±1.4) | Fs (24.8±1.4) | Fs (23.4±1.4) |
| ol/(ol+px) | 0.63±0.03 | 0.59±0.03 | 0.65±0.03 |
| Temp. corrected ol/(ol+px) | 0.63±0.03 | 0.60±0.03 | 0.66±0.03 |

**Figure Captions**

**Figure 1**. Near-IR spectrum of Apophis obtained using the SpeX instrument on NASA IRTF. The spectrum exhibits two absorption bands, one centered at ~0.99 μm and the other centered at ~1.9 μm. These two absorption bands are characteristics of olivine-pyroxene assemblages.

**Figure 2**. Comparison between the NIR spectra of Apophis obtained as part of this work (red), Binzel et al. (2009) (blue), and the spectrum obtained by MITHNEOS (green) file a099942.sp117.txt. All spectra are normalized to unity at ~1.5 μm and are offset vertically for clarity.

**Figure 3**. Band I center vs. BAR for Apophis determined in this work (red diamond). Values extracted from Binzel et al. (2009) and MITHNEOS data are depicted as a blue pentagon and green triangle, respectively. Also shown, the values measured for LL, L, and H ordinary chondrites from Dunn et al. (2010). The polygonal region corresponds to the S(IV) subgroup of Gaffey et al. (1993). The dashed line indicates the location of the olivine-orthopyroxene mixing line of Cloutis et al. (1986).

**Figure 4**. Mol% of fayalite (Fa) vs. ferrosilite (Fs) for Apophis. Values determined in this work are depicted as a red diamond. Calculated chemistries from Binzel et al. (2009) and the MITHNEOS data are shown as a blue pentagon and green triangle, respectively. Measured values for LL (squares), L (triangles), and H (circles) ordinary chondrites from Nakamura et al. (2011) are also included. The error bars in the upper right corner correspond to the uncertainties derived by Dunn et al. (2010), 1.3 mol% for Fa, and 1.4 mol% for Fs. Figure adapted from Nakamura et al. (2011).

**Figure 5.** Color-coded a-posterior probability distribution as computed by the trained naïve Bayes classifier (Blue = H, Green = L and Red = LL) for Figure 4 (Fa vs. Fs). The color gradients highlight the decision boundaries. Binzel et al. (2009) measured value is depicted as a black diamond, while values measured for this work and MITHNEOS data are depicted as pink and yellow diamonds, respectively.

**Figure 6.** Top: Probability distribution for H, L and LL classes as computed by the Naïve Bayes classifier. Bottom: Contour plot of the computed maximum posterior likelihood as function of the Mol % of fayalite (Fa) and Ferrosilite (Fs) as calculated by the Naïve Bayes classifier. For each value of Fa and Fs in the range, the classifier outputs the probability for each of the H, L and LL classes. Here, we selected the maximum likelihood (i.e. the maximum probability out of the three classes) and plot it according to the color code. Binzel et al. (2009) value is depicted as a black circle. Values measured for this work and MITHNEOS data are depicted as pink and yellow circles, respectively.

**Figure 7**. Molar content of Fs vs. ol/(ol+px) ratio for Apophis found in this work (red diamond), along with the values calculated from Binzel et al. (2009) (blue pentagon), and the MITHNEOS data (green triangle). Also shown, measured values for LL, L and H ordinary chondrites from Dunn et al. (2010). Black dashed boxes represent the range of measured values for each ordinary chondrite subgroup. Gray solid boxes correspond to the uncertainties associated to the spectrally-derived values. Figure adapted from Dunn et al. (2010).

**Figure 8**. Color-coded a-posterior probability distribution as computed by the trained naïve Bayes classifier (Blue = H, Green = L and Red = LL) for Figure 7 (Fs vs. ol/(ol+px)). Binzel et al. (2009) measured value is depicted as a black diamond, while values measured for this work and MITHNEOS data are depicted as pink and yellow diamonds, respectively.

**Figure 9**. Contour plot of the computed maximum posterior likelihood as function of ol/(ol+px) and Fs as calculated by the Naïve Bayes classifier. Binzel et al. (2009) value is depicted as a black circle. Values measured for this work and MITHNEOS data are depicted as pink and yellow circles, respectively

**Figure 10.** Plane of sky shape model images of Apophis derived from lightcurve observations by Pravec et al. (2014). (A) shows the orientation of Apophis on January 14, 2013 at 11:07:41 UTC at rotation phase of 189.4°; (B) orientation of Apophis when it was observed by MITHNEOS on January 17, 2013 at 10:59:11 UTC at rotation phase of 5.6°. Due to high degree of uncertainty in the spin state for 2005, we have not included the shape model for those observations. The X indicates the sub-Earth point and the arrow through the shape model is the spin vector.

**Figures**

**Figure 1.**

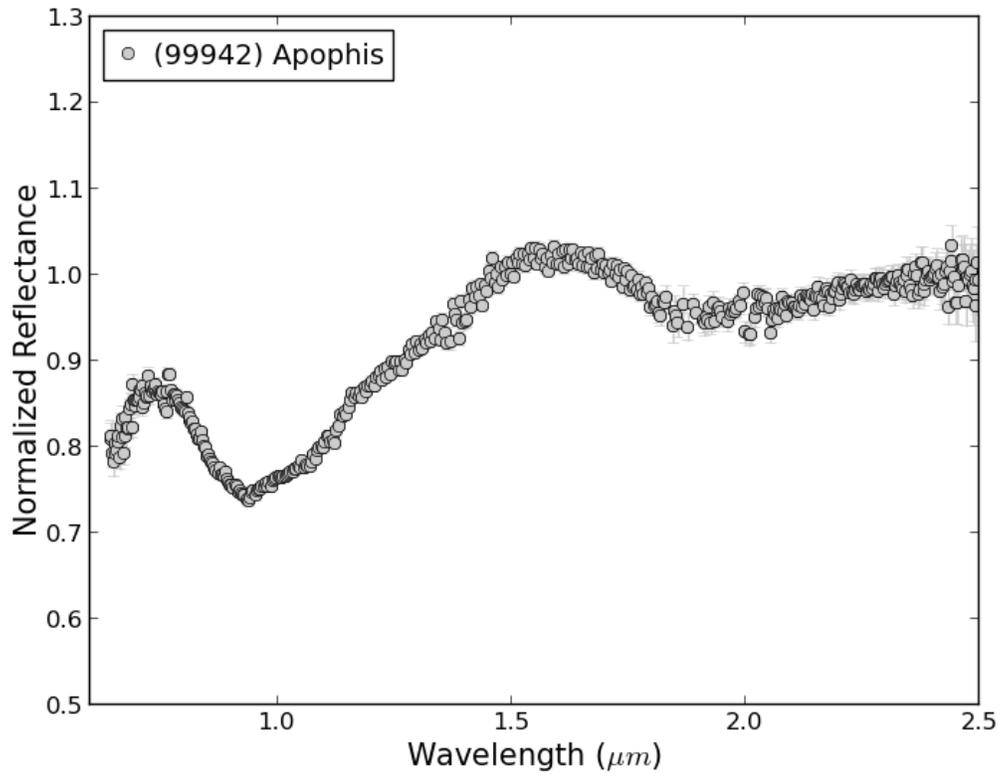

**Figure 2.**

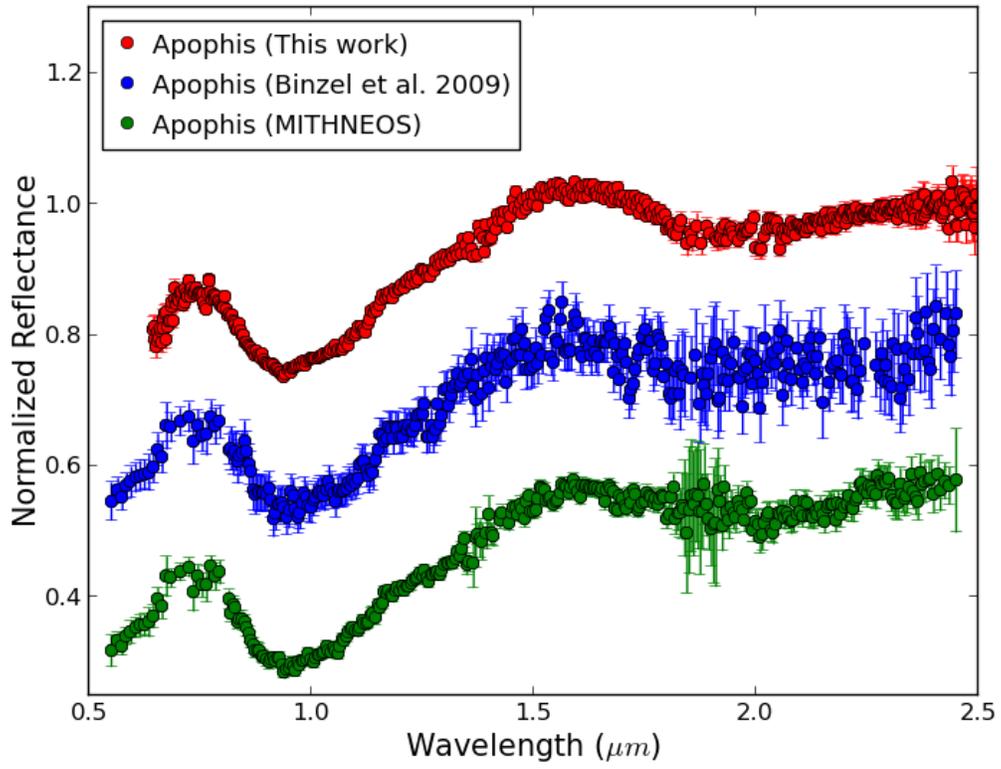

**Figure 3.**

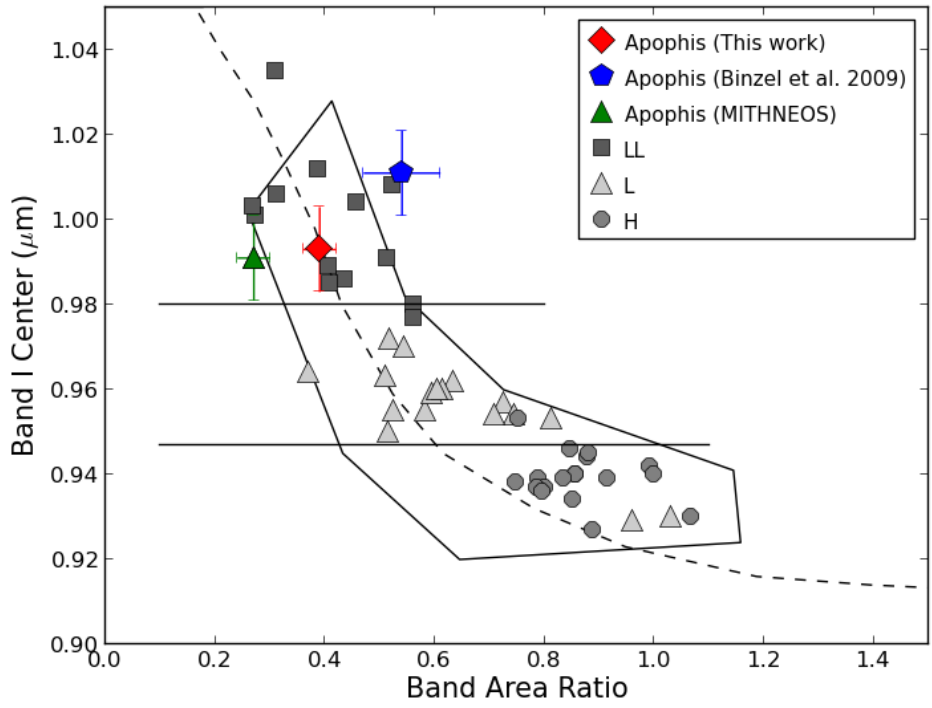

**Figure 4.**

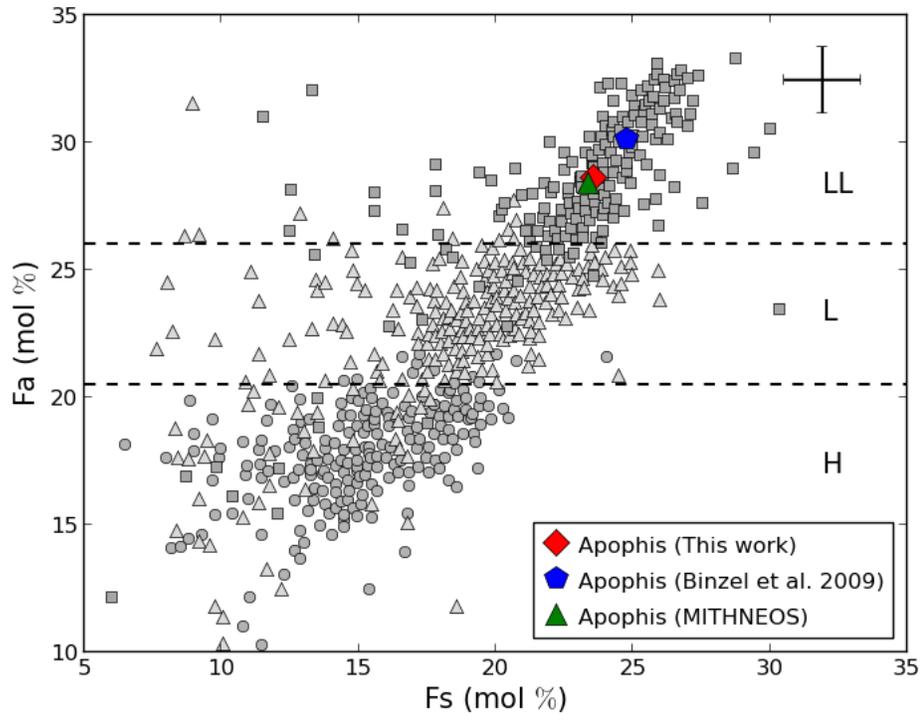
25

**Figure 5.**

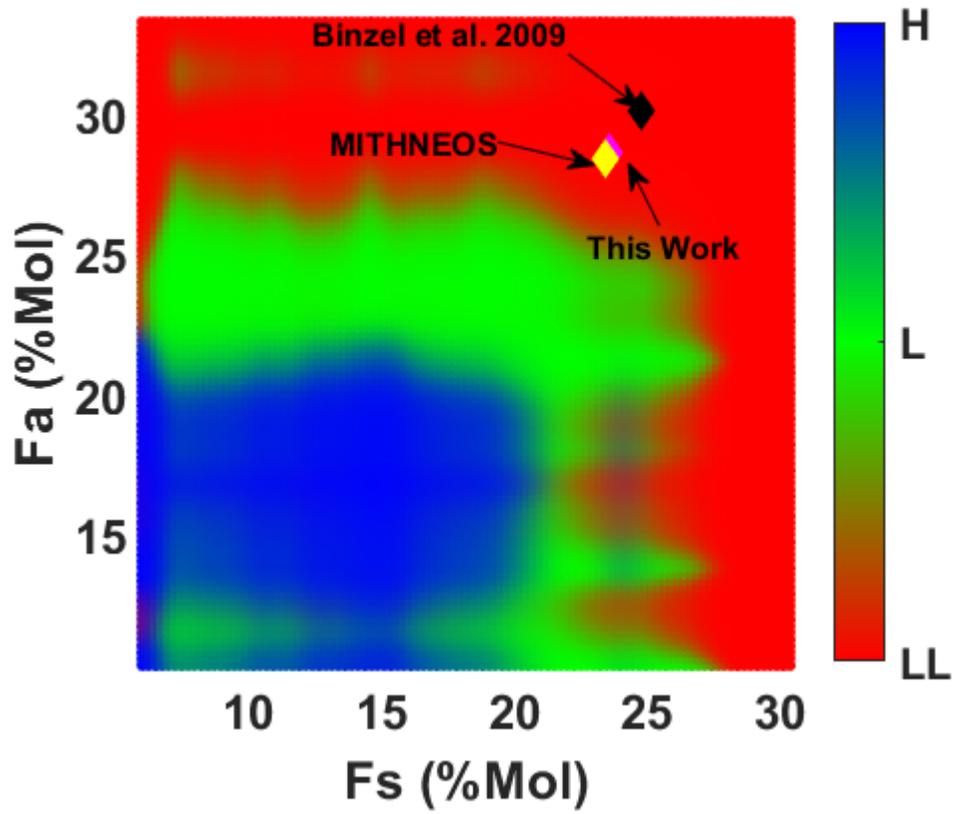

**Figure 6.**

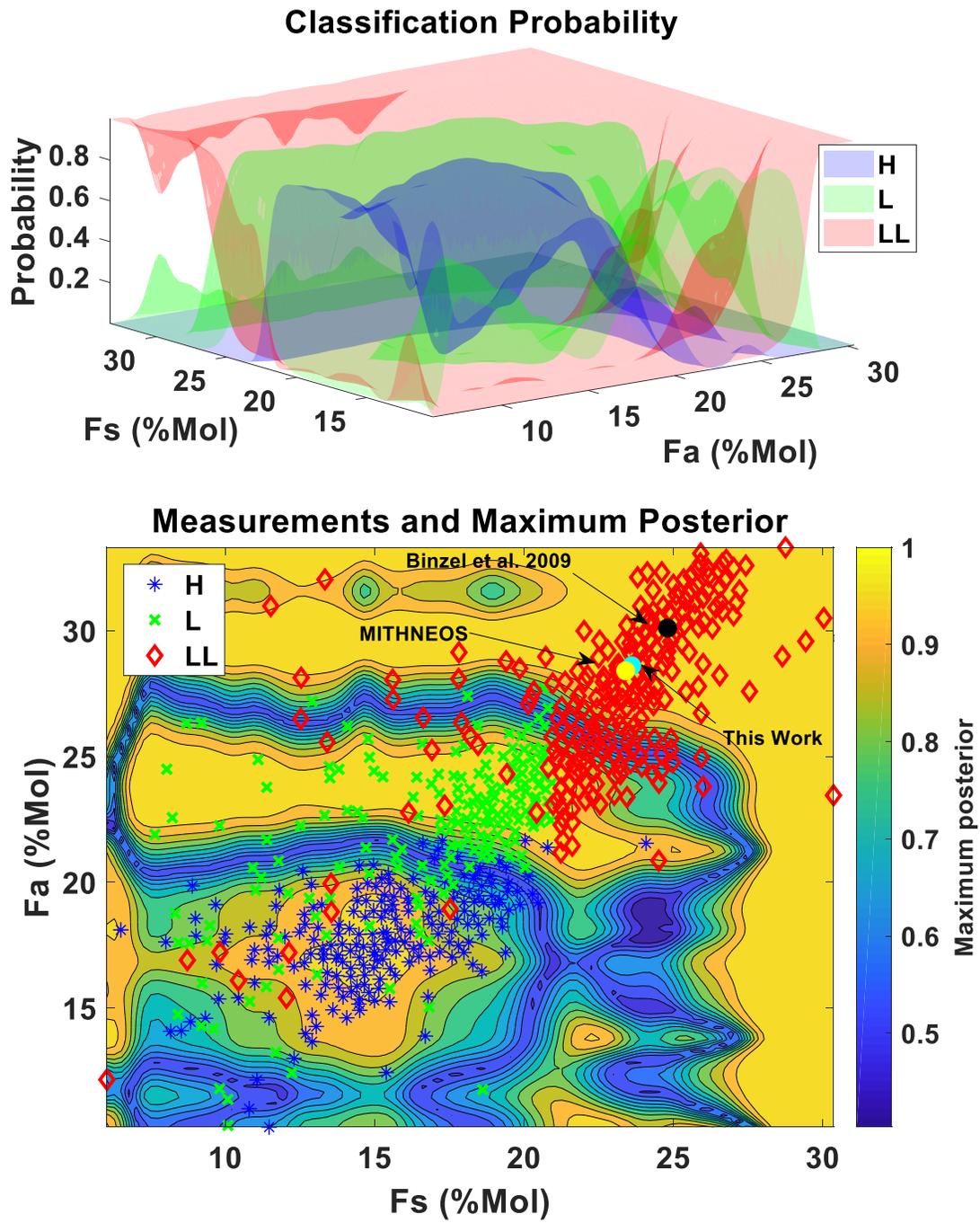

**Figure 7.**

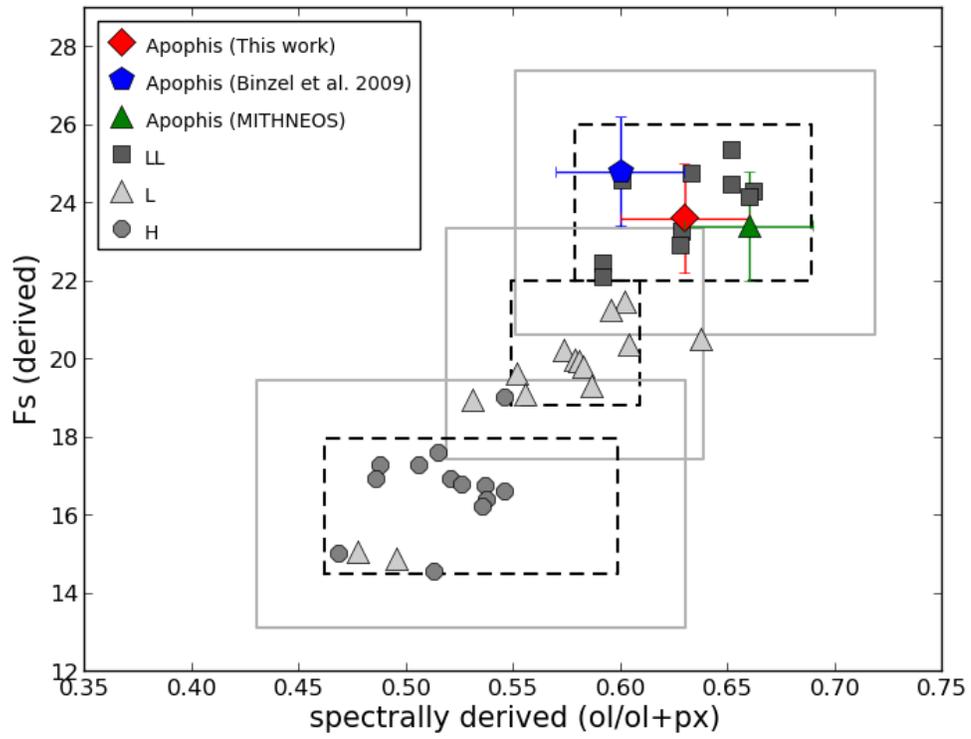

**Figure 8.**

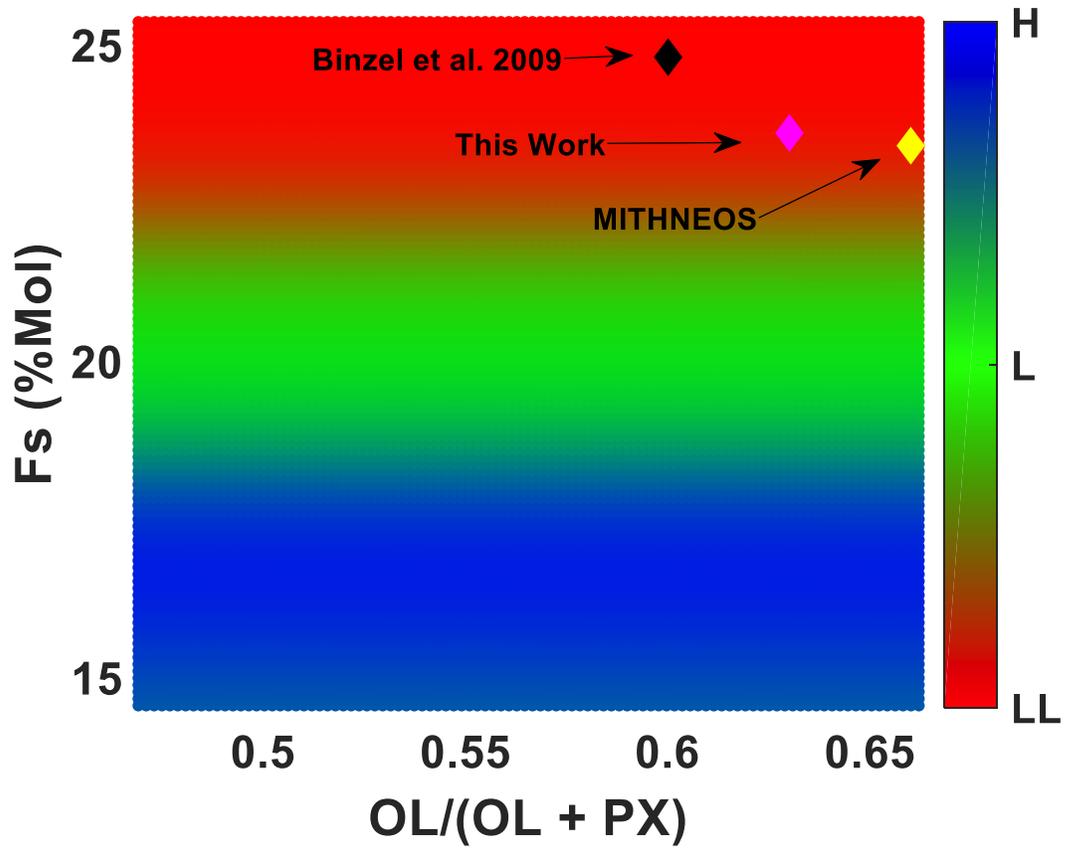

**Figure 9.**

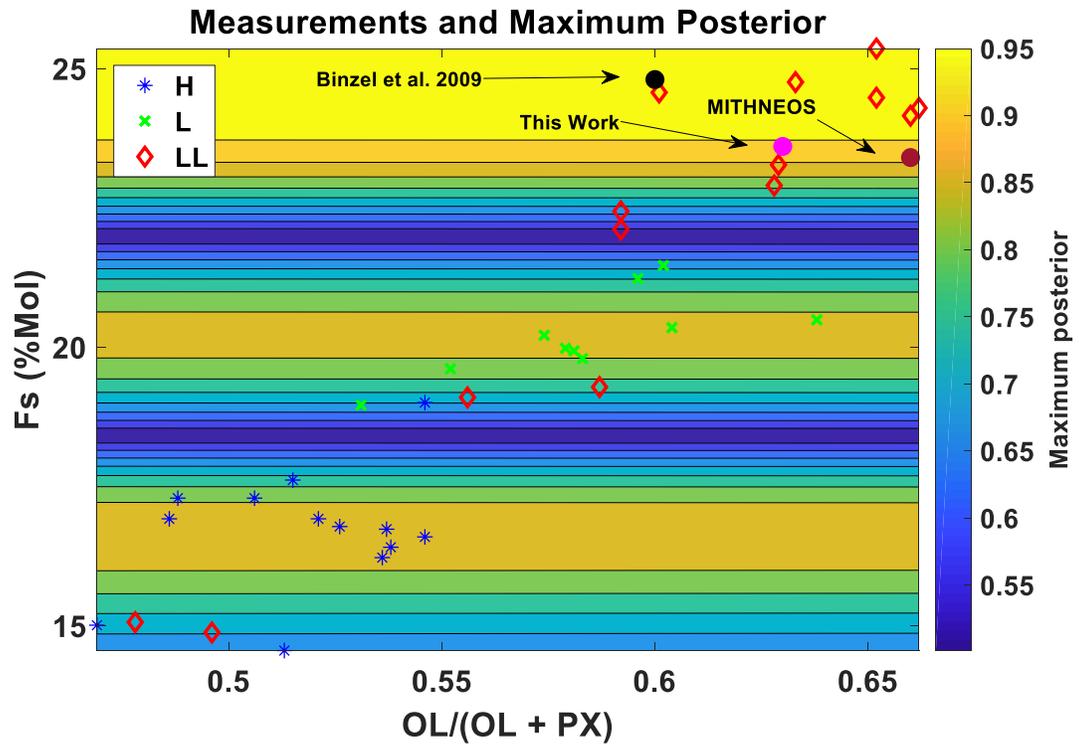

**Figure 10.**

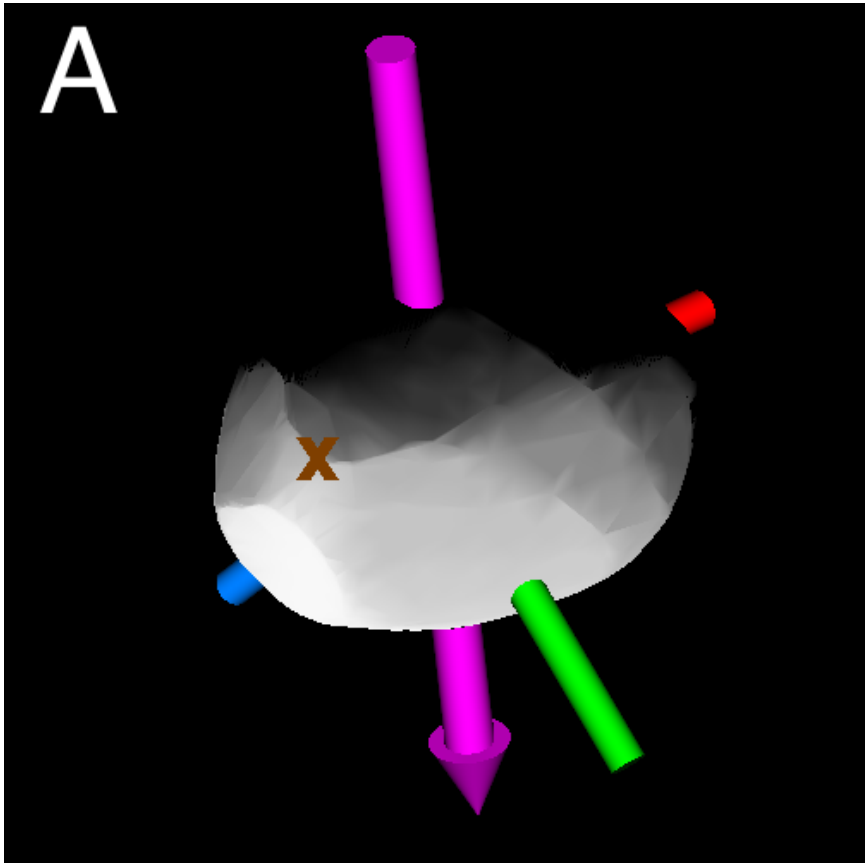

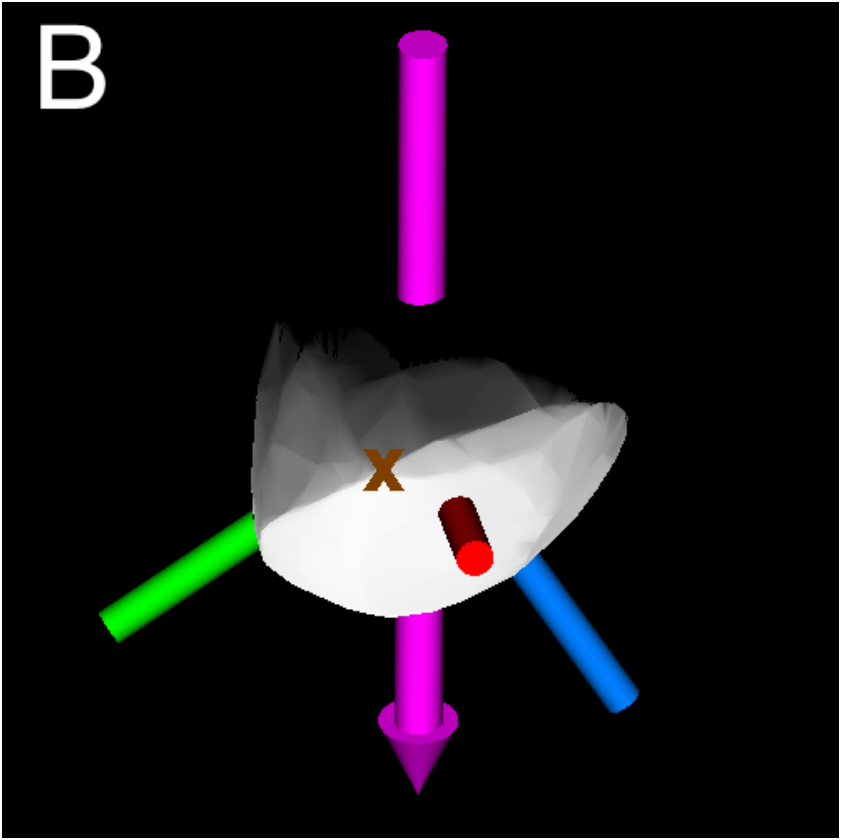